\documentclass[prl,reprint,superscriptaddress,preprintnumbers]{revtex4-2}
\pdfoutput=1
\usepackage{amsfonts, amssymb, amsmath}
\usepackage{mathrsfs}
\usepackage[utf8]{inputenc}
\usepackage[russian,english]{babel}

\usepackage{color}
\definecolor{dark-gray}{gray}{0.20}
\definecolor{gray}{gray}{0.30}
\definecolor{light-gray}{gray}{0.80}
\definecolor{dark-red}{rgb}{0.7,0,0}
\definecolor{dark-green}{rgb}{0.1,0.4,0}
\definecolor{dark-blue}{rgb}{0.3,0.3,0.7}
\definecolor{light-blue}{rgb}{0.8,0.8,1}
\definecolor{blue}{rgb}{0,0,1}
\definecolor{red}{rgb}{1,0,0}
\definecolor{green}{rgb}{0,1,0}

\usepackage{hyperref}
\hypersetup{
	colorlinks=true,
	linkcolor=dark-blue,
	citecolor=dark-red,
	urlcolor=dark-blue,
	linktoc=page
}

\def\cF{{\cal F}}
\def\cI{{\cal I}}

\def\cN{{\cal N}}
\def\cO{{\cal O}}

\def\U{{\rm U}}

\def\i{{\rm i}}

\newcommand{\be}{\begin{equation}}
\newcommand{\ee}{\end{equation}}
\newcommand{\bea}{\begin{eqnarray}}
\newcommand{\eea}{\end{eqnarray}}

\begin{document}

\title{ABJM at finite $N$ via 4d supergravity}

\author{Kiril Hristov}

\affiliation{Faculty of Physics, Sofia University, J. Bourchier Blvd. 5, 1164 Sofia, Bulgaria}
\affiliation{INRNE, Bulgarian Academy of Sciences, Tsarigradsko Chaussee 72, 1784 Sofia, Bulgaria}

\begin{abstract}
We apply the conjecture of \cite{Hristov:2021qsw} for gravitational building blocks to the effective supergravity description of M-theory on S$^7/\mathbb{Z}_k$. Utilizing known localization results for the holographically dual ABJM theory, we determine a complete tower of higher derivative corrections to the AdS$_4$ supergravity and a further set of quantum corrections. This uniquely fixes the gravitational block, leading to holographic predictions for a number of exact ABJM observables, excluding only constant and non-perturbative corrections in the gauge group rank $N$. The predicted S$^3$ partition function is an Airy function that reproduces previous results and generalizes them to include arbitrary squashing and mass deformations/R-charge assignments.  The topologically twisted and superconformal indices are instead products of two different Airy functions, in agreement with direct numeric calculations in the unrefined limit of the former object. The general fixed-point formula for an arbitrary supersymmetric background is similarly given as a product of Airy functions.
\end{abstract}
\date{\today}
\maketitle


\section{Introduction and main results}
\label{sec:intro}
The theory of $N$ coincident M2-branes probing a $\mathbb{C}/\mathbb{Z}_k$ singularity, known as ABJM theory \cite{Aharony:2008ug}, is the most prominent among a general class of 3d Chern-Simons-matter models with holographically dual descriptions. At large $N$ and fixed $k$, ABJM is strongly coupled and one can only rely on supersymmetric localization for the evaluation of protected observables in terms of matrix models. In the same regime of validity, the dual M-theory on asymptotic AdS$_4 \times$S$^7/\mathbb{Z}_k$ is instead weakly coupled and can be compactified to an effective 4d gauged supergravity theory at small Newton constant $G_N$. Many of the M-theory quantum effects manifest themselves in higher derivative terms in the 4d action, but such corrections are not yet well-understood in 11d and in turn cannot be directly derived in 4d. In this work we apply the AdS/CFT dictionary in both directions of the correspondence, leveraging the partial knowledge on one side to fill in missing pieces on the other, and vice versa. First we use known localization results to derive explicitly a full tower of higher derivative corrections to the 4d bulk supergravity and further quantum corrections. In turn we are able to predict exactly a number of ABJM partition functions and related correlation functions, excluding only terms of order $N^0$ and non-perturbative corrections. The reported progress gives significant boost in understanding the strongly interacting 3d $\cN \geq 2$ gauge theories with a gravity dual, and in the same time reveals the structure of quantum gravity corrections.

In more technical detail, our results are a direct application of the conjecture of \cite{Hristov:2021qsw} about gravitational building blocks at full quantum level. It proposes a gluing mechanism in analogy to the 4d $\cN=2$ field theory factorization in terms of Nekrasov partition functions \cite{Nekrasov:2002qd,Nekrasov:2003vi}. The gravitational block serves to determine the bulk on-shell action for various asymptotically AdS$_4$ backgrounds with fixed points of the canonical isometry, as first noticed in \cite{BenettiGenolini:2019jdz} and \cite{Hosseini:2019iad}. Particularly important examples are the asymptotically Euclidean AdS$_4$ backgrounds with squashed S$^3$ boundary slicing, \cite{Martelli:2011fu,Martelli:2011fw}, which exhibit a {\it single} fixed point. This implies a holographic identification of the gravitational building block with the sphere partition function, \cite{Kapustin:2009kz,Hama:2011ea,Imamura:2011wg}. Even though the complete result for ABJM on S$^3$ with arbitrary squashing and mass deformations was not available in literature until now, important results were obtained in various limits, \cite{Fuji:2011km,Marino:2011eh,Nosaka:2015iiw,Hatsuda:2016uqa,Chester:2021gdw}. They are precisely sufficient to {\it holographically bootstrap} the full answer for the gravitational block after superimposing on them some standard supergravity constraints. This allows us to extend the results in \cite{Bobev:2020egg,Bobev:2021oku} obtained by the same logic, determining a complete tower of higher derivative corrections, see \eqref{eq:central}-\eqref{eq:identif}.

Inverting the direction of AdS/CFT, we arrive at the following prediction for the squashed sphere partition function of ABJM theory, 
\be
\label{eq:1}
	Z_{S^3} (b; \Delta_i) \simeq \text{Ai} [C_{S^3}^{-1/3} (N  -B_{S^3})]\ , 
\ee
with the parametrization
\begin{widetext}
\be
	C_{S^3} = \frac{2}{\pi^2 k (b+b^{-1})^4 \Delta_1 \Delta_2 \Delta_3 \Delta_4}\ , \qquad
	B_{S^3} = \frac{k}{24}+  \frac{1}{48 k \Delta_1 \Delta_2 \Delta_3 \Delta_4} \left(k_\mathbb{T} (\Delta) + \frac{(b-b^{-1})^2}{(b+b^{-1})^2}\, k_\mathbb{W} (\Delta) \right)\ ,
\ee
\be
\hspace{-7mm}
\label{eq:k}
k_\mathbb{T} (\Delta) := \sum_i  (\Delta_i)^2 - \frac{(\Delta_1+\Delta_2-\Delta_3 - \Delta_4) (\Delta_1-\Delta_2+\Delta_3-\Delta_4) (\Delta_1 - \Delta_2 - \Delta_3+\Delta_4)}{\sum_i \Delta_i}\ , \quad
k_\mathbb{W} (\Delta) := -2 \sum_{i<j} \Delta_i \Delta_j\ ,
\ee
\end{widetext}
wher $b>0$ is the real geometric squashing, and $\Delta_i$, $i = 1, ..., 4$ are the complexified R-charges (related to the three independent complexified masses as given in \eqref{eq:mass}, see \cite{Jafferis:2010un}), such that $\sum_{i=1}^4 \Delta_i =2$. The ``$\simeq$'' sign in \eqref{eq:1} and below denotes equality upto an overall prefactor of ${\cal O} (N^0)$ that in general depends on $k, b, \Delta_i$, and upto additive terms that constitute non-perturbative corrections. Note that we restrict $|Arg(C_{S^3})| < \pi$ in order not to intersect any Stokes or anti-Stokes lines of the Airy function, securing a unique asymptotic expansion, \eqref{eq:asymp}. A similar form for the squashed sphere partition function was independently proposed in \cite{Bobev:2022jte,Bobev:2022eus}. We observe that the conformal point with canonical R-charges $\bar \Delta_i = \frac12$, $\forall i$, remains unchanged. The above partition function acts as a generating functional for an infinite class of correlations functions that can be computed via derivatives w.r.t. $b$ and $\Delta_i$, with vanishing one-point function (known as F-extremization) and the two-point function of the stress-tensor fixed by the coefficient $c_T$ as in \eqref{eq:ct}.

We also apply the gluing rules \cite{Hosseini:2019iad,Hristov:2021qsw} for rotating black hole spacetimes in AdS$_4$, which exhibit two different fixed points. The two topologically distinct classes of black holes preserving supersymmetry with (\cite{Cacciatori:2009iz,Hristov:2018spe}), or without (\cite{Hristov:2019mqp}), a twist correspond respectively to the {\it refined} topologically twisted \cite{Benini:2015noa,Benini:2015eyy} and superconformal \cite{Kim:2009wb,Imamura:2011su,Kapustin:2011jm} indices (rTTI and SCI) of ABJM theory. We present our holographic predictions for the rTTI and SCI, which both depend on a set of real magnetic charges $\frak{n}_i$ and complex chemical potentials $\Delta_i$, $\omega$ conjugate respectively to the conserved real electric charges $\frak{q}_i$ and angular momentum $\frak{J}$. They are given by, \footnote{These results assume that no other supergravity saddles contribute to the path integral at the same orders in $N$, in particular that no multicentered black holes exist with the same boundary conditions.}
\bea
\label{eq:tti}
\begin{split}
	Z_\text{rTTI} (\frak{n}_i, \omega, \Delta_i) \simeq & \text{Ai} [C_+^{-1/3} (N  -B^0_+)] \\ &\times \text{Bi} [C_-^{-1/3} (N  -B^1_-)]\ , 
\end{split}
\eea
with $\sum_i \frak{n}_i = 2$ and $\sum_i \Delta_i = 2$; and
\bea
\label{eq:sci}
\begin{split}
	Z_\text{SCI} (\frak{n}_i, \omega, \Delta_i) \simeq & \text{Ai} [C_+^{-1/3} (N  -B^0_+)]\, \\
 & \times \text{Ai} [C_-^{-1/3} (N  -B^0_-)]\ , 
\end{split}
\eea
with $\sum_i \frak{n}_i = 0$ and $\sum_i \Delta_i = 2 (1+\omega)$; where
\begin{widetext}
\be
\label{eq:BC}
	C_\pm = \frac{2\, \omega^2}{\pi^2 k \prod_i (\Delta_i \pm \omega \frak{n}_i)}\ , \quad
	B^s_\pm = \frac{k}{24}+  \frac{(\omega+(-1)^s)^2 k_\mathbb{T} (\Delta \pm \omega \frak{n}_i) + (\omega- (-1)^s)^2 k_\mathbb{W} (\Delta \pm \omega \frak{n}_i) }{48 k  \prod_i (\Delta_i \pm \omega \frak{n}_i)}  \ ,
\ee
\end{widetext}
with $k_\mathbb{T, W}$ defined in \eqref{eq:k}. We again restrict $|Arg (C_\pm)| < \pi $ in both cases, and further assume $\text{csgn} (\sqrt{\prod_i (\Delta_i \pm \omega \frak{n}_i)}/\omega) = 1$ in order to fix the orientation of the fixed points (see \eqref{eq:qgravNek} and \eqref{eq:csgn}) - a reversal of the sign results in a flip between $\text{Ai}$ and $\text{Bi}$ in the formulae above. The rTTI and SCI can also be considered as generating functionals for an infinite class of correlation functions with non-vanishing one-point functions (knows as $\cI$-extremization), see \eqref{eq:Iextrem}. The twisted branch of black holes further admits a static limit, corresponding to the usual (unrefined) TTI: $Z_\text{TTI} = Z_\text{rTTI} (\omega = 0)$. Our supergravity prediction in this case agrees {\it precisely} with a recent large $N$ numerical evaluation \cite{Bobev:2022jte} of the matrix model \cite{Benini:2015eyy}, see \eqref{eq:static}. This constitutes a very strong confirmation of the factorization \cite{Hristov:2021qsw} as it successfully predicts an {\it infinite} set of subleading corrections.

\section{4d supergravity}
\label{sec:setup}
As discussed at length in \cite{Hristov:2021qsw}, whose notation and conventions we follow here, the relevant Lagrangian of gauged higher derivative 4d $\cN=2$ supergravity in the presence of $n_V$ physical abelian vector multiplets and no physical hypermultiplets is specified by the prepotential
\be
\label{eq:prepot}
	F(X^I; A_\mathbb{W},  A_\mathbb{T}) = \sum_{m, n = 0}^\infty \, F^{(m,n)} (X^I)\, (A_\mathbb{W})^m\, (A_\mathbb{T})^n\ ,  
\ee
together with the choice of gauging parameters, $g_I$. The above formula features the (off-shell) vector multiplet complex scalars $X^I$, $I = 0, ... , n_V$, and the composite scalars $A_\mathbb{W},  A_\mathbb{T}$ responsible for the higher derivative {\it Weyl-squared} \cite{Bergshoeff:1980is} and {\it T-log} \cite{Butter:2013lta} invariants, respectively. Supersymmetry constrains the holomorphic functions $F^{(m,n)} (X^I)$ to be homogeneous of degree $2 (1-m-n)$ giving rise to terms with $2 (1+m+n)$ derivatives, each of them suppressed by an additional power of the Newton constant, starting with the two-derivative Lagrangian proportional to $1/G_N$. 

Here we are interested in the effective 4d supergravity description of 11d supergravity compactified on S$^7 / \mathbb{Z}_k$ for finite $k$, truncated to its $\U(1)^4$-invariant subsector. At two derivatives we have the STU model, $n_V = 3$, with a standard choice for the gauging parameters $g_I = 1, \forall I$ and $F^{(0,0)} = -2 \i \sqrt{X^0 X^1 X^2 X^3}$, \cite{Cvetic:1999xp}, thus setting the AdS$_4$ length scale to $L=1/\sqrt{2}$. The holographic dictionary at two derivatives gives $1/(4 G_N) = \sqrt{2 k}/3\, N^{3/2}$. One of our central results (we postpone the derivation to the next section) is that the full 4d higher derivative supergravity Lagrangian is specified by
\bea
\label{eq:central}
\begin{split}
	F =& - 2\i \sqrt{X^0 X^1 X^2 X^3}\, \\ &\times \sum_{n=0}^\infty f_n \left( \frac{ k_{\mathbb{W}} (X) A_{\mathbb{W}} + k_{\mathbb{T}} (X) A_{\mathbb{T}}}{64\, X^0 X^1 X^2 X^3} \right)^n\ ,
\end{split}
\eea
with the functions $k_{\mathbb{W}, \mathbb{T}}$ given in \eqref{eq:k}, constrained by supersymmetry to be homogeneous of second degree. The constant coefficients $f_n$ in this infinite expansion can be holographically identified as
\bea
\label{eq:identif}
\begin{split}
	\frac{f_0}{4 G_N} &= \frac{\sqrt{2 k}}{3} \left( N^{3/2} - \frac1{16}\, k N^{1/2} + ...  \right) = \frac{\sqrt{2 k}}{3}\, N_k^{3/2} , \\
	2 \pi f_1 &= \left(-\frac1{2 k} \right)\,  \frac{\sqrt{2 k}}{3}\, N_k^{1/2} , \\
& ...  \\
\frac{2 \pi f_n}{\kappa^{2 (1-n)}} &= \left( \frac{(2 n-5)!!\, 3 }{n!\, (6 k)^n}   \right)\, \frac{\sqrt{2 k}}{3}\, N_k^{3/2-n} ,
\end{split}
\eea
where $\kappa^2 := 8 \pi G_N$, and for brevity we introduced the shifted rank
\be
	N_k := N - \frac{k}{24}\ ,
\ee
which is still assumed large in the supergravity approximation. Notice that the first terms on the r.h.s. of \eqref{eq:identif} can be seen to originate from the Taylor expansion of $(1-1/(3k))^{3/2}$ around small $1/(3 k)$, such that we can resum \eqref{eq:central} to
\bea
	\label{eq:central-resum}
\begin{split}
	F =& - \frac{\i \kappa^2\, \sqrt{2 k X^0 X^1 X^2 X^3}}{3 \pi}\, \\ & \times  \left(N_k -\frac{1}{3 k}\, \frac{ ( k_{\mathbb{W}} (X) A_{\mathbb{W}} + k_{\mathbb{T}} (X) A_{\mathbb{T}})}{64 \kappa^2\, X^0 X^1 X^2 X^3} \right)^{3/2}\ .
\end{split}
\eea
We still have $g_I = 1$, $\forall I$, and without loss of generality can set $f_0 = 1$ as done in the discussion above \eqref{eq:central}. We should emphasize here that some of the results we have used in obtaining \eqref{eq:central}-\eqref{eq:identif} hold only at $k=1, 2$, but it is tempting to speculate that the expressions are meaningful all the way to the type IIA regime where $k \sim N$ (keeping $N_k$ large); in this case each term in the supergravity expansion comes with an extra suppression of $1/N^2$ as expected.

\vspace{-5mm}
\subsection{Gravitational building blocks}
\label{sec:Nek}
\vspace{-3mm}

{\bf Higher derivatives.} According to \cite{Hristov:2021qsw}, the on-shell action of any supersymmetric background with fixed points $\sigma$ is given by
\be
\hspace{-4mm}
\label{eq:onshell}
	\cF (\chi^I, \omega) = 4 \i \pi^2\, \sum_\sigma \frac{F (\kappa^{-1} X^I_\sigma (\chi^I, \omega); (1-\omega_\sigma)^2, (1+\omega_\sigma)^2)}{s_\sigma \omega_{\sigma}}\ ,
\ee
such that $\cF =: \sum_\sigma \cF_\sigma$, where $\chi^I$ are the so-called Coulomb branch parameters and $\omega$ is the geometric deformation parameter. The precise identification of the signs $s_\sigma$ and complex parameters $X^I_\sigma, \omega_\sigma$ at each fixed point, together with one overall constraint on the parameters $\lambda (g_I, \chi^I, \omega)$, is determined specifically for each background and goes under the name {\it gluing rule}, \cite{Hosseini:2019iad}. \cite{Hristov:2021qsw} then defined the {\it gravitational block} as the partition function corresponding to a single fixed point contribution, $- \log Z^\text{sugra}_\sigma = \cF_\sigma$. Using the entire higher derivative prepotential \eqref{eq:central} and the holographic dictionary \eqref{eq:identif}, it is straightforward to obtain
\bea
\label{eq:gravNek}
\nonumber	Z^\text{sugra}_\sigma &=& \exp \Big[ -\frac{4 \pi \sqrt{2 k X^0_\sigma X^1_\sigma X^2_\sigma X^3_\sigma}}{3 s\, \omega} \left( N_k - B_\sigma \right)^{3/2} \Big]\ ,  \\  B_\sigma&: = & \frac{k_\mathbb{T} (X_\sigma) (1+\omega)^2 + k_\mathbb{W} (X_\sigma) (1-\omega)^2}{192 k X^0_\sigma X^1_\sigma X^2_\sigma X^3_\sigma}\, .
\eea

{\bf UV completion.} The gravitational block \eqref{eq:gravNek} captures all the information from higher derivative supergravity, but has no knowledge of the full UV completion within string/M-theory. For the model at hand we can make this precise by once again invoking AdS/CFT and observing that the sphere partition function of ABJM takes the form of an Airy function. The asymptotic expansions of the Airy functions of first and second kind read
\bea
\label{eq:asymp}
\begin{split}
\text{Ai} (z) &\sim \frac{e^{-2/3\, z^{3/2}}}{2 \sqrt{\pi} z^{1/4}}\, \Big[ \sum_{n=0}^\infty \frac{(-1)^n 3^n \Gamma(n+\frac56) \Gamma(n+\frac16)}{2 \pi n! 4^n z^{3n/2} } \Big]  \ , \\
\text{Bi} (z) &\sim \frac{e^{+2/3\, z^{3/2}}}{\sqrt{\pi} z^{1/4}}\, \Big[ \sum_{n=0}^\infty \frac{3^n \Gamma(n+\frac56) \Gamma(n+\frac16)}{2 \pi n! 4^n z^{3n/2} } \Big] \ ,
\end{split}
\eea
in the region $|Arg(z)| < \pi/3$. Comparing with \eqref{eq:gravNek}, we can identify the gravitational block only with the leading exponential term of the asymptotic expansions. The parameter space of $\chi^I, \omega$ on the underlying background serves to determine the overall sign of the exponent, given by $\text{csgn}$ of the argument of \eqref{eq:gravNek}. We can naturally split the set of fixed points $\sigma$ on a given supersymmetric background into positive and negative orientations, $\sigma_+$ and $\sigma_-$, in line with the expansion of $\text{Ai}$ and $\text{Bi}$, respectively. 

Clearly, the higher derivative corrections are not enough to capture the complete asymptotic expansions and one requires output from the UV completion, also subject of fixed-point factorization according to \cite{Hristov:2021qsw}, \footnote{This logic can be alternatively confirmed via supergravity localization, \cite{Dabholkar:2010uh,Hristov:2018lod}, observing that the one-loop corrections to the on-shell action from each fixed point give rise precisely to the type of contributions in the denominator of \eqref{eq:asymp}, see \cite{Hristov:2021zai}.}. Such additional corrections come from the entire Kaluza-Klein tower of massive modes that are not captured by the effective 4d Lagrangian. Deriving the entire set of corrections in \eqref{eq:asymp}, including the subleading sums in the brackets, is outside the present scope.  Instead we can {\it postulate} via AdS/CFT the corresponding quantum correction to the higher derivative part of our result, in line with \cite{Hristov:2021qsw}. This is possible in a unique fashion since we can reconstruct the type of the Airy function from the exponentials in \eqref{eq:asymp}. The UV completed {\it quantum gravitational blocks} then take the following form,
\bea
\label{eq:qgravNek}
\hspace{-3mm}
\begin{split}
 Z_{\sigma_+} =& \text{Ai} (C_{\sigma}^{-1/3} (N_k - B_{\sigma}) )\ , \, \, \, Z_{\sigma_-}= \text{Bi} (C_{\sigma}^{-1/3} (N_k - B_{\sigma}) )\ , \\
	 C_\sigma :=& \frac{\omega^2}{8 \pi^2 k\, X^0_\sigma X^1_\sigma X^2_\sigma X^3_\sigma}\ , \,  \, \, \text{sgn}_\sigma = \text{csgn} \left( \frac{\sqrt{\prod_I X^I_\sigma}}{s\, \omega} \right)\ .
\end{split}
\eea
The general prediction for a full quantum gravitational partition function of supersymmetric backgrounds with AdS$_4 \times$S$^7 / \mathbb{Z}_k$ asymptotics therefore becomes a product over Airy functions,
\be
	Z  (\chi^I, \omega) = \prod_{\sigma_+}  Z_{\sigma_+} \, \prod_{\sigma_-} Z_{\sigma_-}\ ,
\ee
and depends on the correct identification of the gluing rules for each background. We now turn to discuss explicitly the three main examples mentioned in the introduction.

\section{Squashed sphere partition function}
\label{sec:sphere}

The holographic squashed S$^3$ background is our main example, as it exhibits only a single fixed point in the centre of the geometry, \cite{BenettiGenolini:2019jdz}. Applying the gluing rule proposed in \cite{Hristov:2021qsw} on this single fixed point, we have the relations $s = 1, X^I = (1+\omega)\, \chi^I$ and $\omega = b^2$ under the constraint $g_I \chi^I = \sum_I \chi^I = 1$. Since the deformation $\omega$ in this case is related to the real squashing parameter $b$, the fixed point is of positive orientation by the definition above, \footnote{Note that the background superimposing the squashings of \cite{Martelli:2011fu,Martelli:2011fw} with the arbitrary R-charge assignments of $\chi^I$ \cite{Freedman:2013oja} has not yet been constructed. The determination is simply fixed by the fact that $b$ is positive.}. In order to convert to the standard field theory notation, we further need to identify $\chi^I =\frac12\, \Delta_i$,  \footnote{By a standard abuse of notation due to the well-established conventions on both sides of the duality, we need to identify the upper symplectic indices $I = 0, ..., 3$ with the flat lower indices $i = 1, ..., 4$. The precise map is irrelevant since all quantities on both sides are permutation-invariant.}, leading us to the anticipated final answer in \eqref{eq:1}-\eqref{eq:k}. It is in agreement with the large $N$ evaluation of \cite{Choi:2019dfu,Hosseini:2019and} and a number of finite $N$ limits, see below.

The one-point correlation functions need to vanish at the conformal point,
\be
	\partial_b Z_{S^3}|_{\bar b, \bar \Delta} = \partial_{\Delta_i} Z_{S^3}|_{\bar b, \bar \Delta} = 0\ ,
\ee
which is the concept of F-extremization \cite{Jafferis:2010un,Jafferis:2011zi}. The conformal point is well-known to be
\be
	\bar b = 1 , \qquad \bar \Delta_i = \frac12\ , \forall i\ ,
\ee
corresponding holographically to the maximally symmetric AdS$_4$ space with round S$^3$ boundary and no additional deformations that would trigger an RG flow. One can similarly compute the coefficient of the two-point stress-tensor correlator, \cite{Closset:2012ru}
\be
\label{eq:ct}
	c_T = - \frac{32}{\pi} \text{Re}\,  \partial^2_b \log Z_{S^3} |_{b=1, \bar \Delta}\ ,
\ee
and various other correlation functions, see \cite{Chester:2020jay} for more details. The fact that we have now determined the complete dependence on the squashing $b$, apart from the $N^0$ term, means that one can evaluate these correlation functions at much higher precision than previously known.

We now turn to a discussion of several special limits of the partition function and a two step explanation of how our main results \eqref{eq:central}-\eqref{eq:identif} were derived.

\vspace{-5mm}
\subsection{Special limits and bootstrapping $F(X^I; A_\mathbb{W},  A_\mathbb{T})$}
\label{sec:boot}
\vspace{-3mm}

{\bf Round sphere, step 1.}
Let us first consider the case $ b=1$. In the supergravity evaluation this corresponds to the Nekrasov-Shatashvili limit, $A_\mathbb{W} = 0$ in \eqref{eq:prepot}. Two important analytic results are available via a direct matrix model evaluation of the round S$^3$ partition function. At the conformal point, \cite{Fuji:2011km,Marino:2011eh} derived the Airy function \eqref{eq:1} with
\be
	C_{S^3} (1; \bar \Delta) = \frac2{\pi^2 k}\ , \quad B_{S^3} (1; \bar \Delta) =\frac{k}{24}+ \frac1{3 k}\ .
\ee
On the other hand, in the presence of two of the three free R-charges ($\Delta_{i+2} = 1-\Delta_i$ or permutations), \cite{Nosaka:2015iiw} found
\be
	C_{S^3} (1; \Delta) = \frac1{8 \pi^2 k \prod_i \Delta_i}\ , \quad B_{S^3} (1; \Delta) =\frac{k}{24}+ \frac{\sum_i (\Delta_i)^2}{48 k \prod_i \Delta_i}\ .
\ee
In order to derive \eqref{eq:central} and \eqref{eq:identif}, the {\it first step} is to use these two results and compare them with a single gravitational block from an arbitrary prepotential, \eqref{eq:onshell}. Such a relation restricts greatly the prepotential and puts it into the form of \eqref{eq:central} with the holographic identification \eqref{eq:identif}, leaving for the moment $k_\mathbb{W}$ arbitrary and fixing only the first term of $k_\mathbb{T}$ in \eqref{eq:k}. At four derivatives, this part of $k_\mathbb{T}$ was successfully identified already in \cite{Bobev:2021oku} based on the same logic.

{\bf Squashed sphere, step 2.}
In the case of canonical R-charge assignments and arbitrary squashing, we find
\be
\hspace{-2mm}
	C_{S^3} (b; \bar \Delta) = \frac{32}{\pi^2 k (b+b^{-1})^4}\ ,  B_{S^3} (1; \bar \Delta) =\frac{k}{24}+ \frac{1 - 3 \frac{(b-b^{-1})^2}{(b+b^{-1})^2}}{3 k }\ .
\ee
This is a new prediction that has not been derived in the literature from a direct matrix model calculation. However, we can successfully compare it with the known special case $b=\sqrt{3}$ in \cite{Hatsuda:2016uqa} (valid at $k=1$ there)
\be
	C_{S^3} (\sqrt{3}; \bar \Delta) = \frac{9}{8 \pi^2 k}\ , \quad B_{S^3} (\sqrt{3}; \bar \Delta) =\frac{k}{24}+ \frac1{12 k}\ .
\ee
In terms of fixing the supergravity prepotential, this result is telling us that $k_\mathbb{W} (\bar \Delta) = -3$, see \cite{Bobev:2020egg}. Due to the restriction on $k_\mathbb{W}$ to be homogeneous function of second degree, one can intuitively understand why it takes the form of \eqref{eq:k}. However, the {\it second step} in deriving \eqref{eq:central} and \eqref{eq:identif} is a crucial matrix model identity derived in \cite{Chester:2021gdw},
\bea
\begin{split}
	Z_{S^3} (b&;  m_1, m_2, \i \frac{b-b^{-1}}{2}) = \\  &Z_{S^3} (1; \frac{b^{-1} m_+ + b\, m_-}2, \frac{b^{-1} m_+ - b\, m_-}2, 0)\ ,  
\end{split}
\eea
valid at $k=1,2$; with $m_\pm := (m_1 \pm m_2) |_{b=1}$, and the mass deformation parameters $m_{1,2,3}$ given by
\be
\label{eq:mass}
	m_1 = \i \frac{(b+b^{-1})}2\, (\Delta_2 + \Delta_3 - 1)\ ,
\ee
and permutations. The above relation can be seen to hold exactly for the result presented in \eqref{eq:1}. It leads to a very strong condition on the form of the functions $k_\mathbb{T}$ and $k_\mathbb{W}$, fixings them uniquely upon the additional constraint on their degree and the partially fixed form from the first step above. This concludes the holographic bootstrap procedure used to derive \eqref{eq:central}-\eqref{eq:identif}, and in turn the rest of our results.

{\bf Cardy limit.} Another interesting limit is the maximal deformation of the three-sphere. For definiteness we take $b \rightarrow 0$ (there is a symmetry under $b \leftrightarrow b^{-1}$), leading to
\bea
\label{eq:cardy}
\begin{split}
	-\log Z_{S^3}^\text{Cardy} &\simeq \frac{\pi \sqrt{2 k \Delta_1 \Delta_2 \Delta_3 \Delta_4}}{3 b^2}\, N_{k, \Delta}^{3/2} + \frac14 \log  N_{k, \Delta}\ , \\
	N_{k, \Delta}& := N - \frac{k}{24} + \frac{\sum_i (\Delta_i)^{-1}}{12 k}\ ,
\end{split}
\eea
where we also kept the $\cO(b^0)$ correction.

\section{Topologically twisted index}
\label{sec:tti}
Let us consider the static \cite{Cacciatori:2009iz,Halmagyi:2014qza} and rotating \cite{Hristov:2018spe} black holes in AdS$_4$ preserving supersymmetry with a twist, with magnetic charges $p^I$ obeying $\sum_I p^I = -1$. In the general rotating case the background exhibits two fixed points situated at the near-horizon geometry AdS$_2 \times_w S^2$: the center of AdS$_2$ and the two poles of the sphere. The gluing rule  was determined in \cite{Hosseini:2019iad}: at the first fixed point we find an arbitrary complex deformation $\omega$, $s=1$ and $X^I = \chi^I - \omega p^I$; the second fixed point exhibits the opposite deformation, $-\omega$, along with $s=1$ and $X^I = \chi^I + \omega p^I$, with the overall constraint $g_I \chi^I = \sum_I \chi^I = 1$. Using the field theory parametrization, $\chi^I = \frac12 \Delta_i$, $p^I = - \frac12 \frak{n}_i$, we arrive at the anticipated result \eqref{eq:tti} with \eqref{eq:BC}.

The determination of the orientation of the fixed points in this case is much more complicated due to the large parameter space including the magnetic charges, \footnote{Similarly to the holographic squashed sphere case, the complete set of Euclidean solutions in the presence of rotation has not been written down. The related static Euclidean saddles with higher genus topology were found in \cite{Bobev:2020pjk}.}. We have presented \eqref{eq:tti} under the assumption that
\be
\label{eq:csgn}
	\text{csgn} (\frac{\sqrt{\prod_i (\Delta_i \pm \omega \frak{n}_i)}}{\omega}) = 1\ ,
\ee
which might not always be the case. Upon flip to $-1$ the given fixed point changes orientation and we need to change $\text{Ai}$ to $\text{Bi}$ in \eqref{eq:tti}, and vice versa, see \eqref{eq:qgravNek}.

The one-point functions at the conformal point are no longer vanishing, but instead relate to the conserved electric charges $\frak{q}$ \footnote{The electric charges are also a subject of a constraint inherited from the constraint on $\Delta_i$, see \cite{Benini:2016rke}.} and angular momentum $\frak{J}$, \cite{Benini:2016rke}
\be
\label{eq:Iextrem}
	\partial_{\Delta_i} \log Z_\text{rTTI}|_{\bar \omega, \bar \Delta} = \i\, \frak{q}_i \ , \quad \partial_\omega \log Z_\text{rTTI}|_{\bar \omega, \bar \Delta} = \i\, \frak{J}\ ,
\ee
which is the concept of $\cI$-extremization \cite{Benini:2015eyy,Benini:2016rke}. Due to the complicated equations we can no longer present a general expression for $\bar \omega$ and $\bar \Delta$, which do receive corrections to their large $N$ values \cite{Benini:2015eyy}. Higher order correlation functions of the flavor charge operators follow by applying additional derivatives. 

\vspace{-5mm}
\subsection{Static/unrefined limit}
\label{sec:static}
\vspace{-3mm}

The usual (unrefined) TTI, recently evaluated directly with a high precision, corresponds to $\omega = 0$ (vanishing rotation). Note that the $1/\omega$ terms in the two Airy functions in \eqref{eq:tti} cancel out exactly, such that the leading term is precisely constant in $\omega$, giving
\bea
\hspace{0mm}
\begin{split}
\nonumber
	 - \log  Z_\text{TTI} &= -\log Z_\text{rTTI} (\omega = 0) \simeq \\ 
 \frac{\pi \sqrt{2 k \prod_i \Delta_i}}{3}\, & \left( \sum_i \frac{\frak{n}_i}{\Delta_i} ( N_{k, \Delta} - k_i )\right)  N_{k, \Delta}^{1/2}+ \frac12 \log  N_{k, \Delta}\ ,
\end{split}
\eea
\vspace{-5mm}
\be
\label{eq:static}
	  k_i := \frac{(2-\Delta_i) \prod_{j \neq i} (\Delta_i+\Delta_j)}{8 k \Delta_1 \Delta_2 \Delta_3 \Delta_4} \ ,
\ee
with $\sum_i \Delta_i = 2$ and $ N_{k, \Delta}$ defined in \eqref{eq:cardy}, which again does not include possible constant and non-perturbative pieces in $N$. The first term is the full contribution from the gravitational building blocks in \eqref{eq:gravNek} and captures the complete higher derivative prepotential \eqref{eq:central}, while only the $\log$ comes from the UV completion in \eqref{eq:qgravNek} as the rest of the subleading terms are proportional to $\omega^{2 n} N_{k, \Delta}^{-3 n}$, $n \in \mathbb{Z}_+$, and vanish at $\omega = 0$. \eqref{eq:static} matches exactly with the remarkably precise numerical evaluation of \cite{Bobev:2022jte} that superseded the results of \cite{Liu:2017vll}. This is a non-trivial check on the factorization proposal of \cite{Hristov:2021qsw}, which can be seen to hold precisely in higher derivative supergravity. To the best of the author's knowledge, this constitutes the most detailed check of the AdS$_4$/CFT$_3$ correspondence to date, matching an infinite set of subleading corrections in the small $G_N$/large $N$ limit.

\section{Superconformal index}
\label{sec:sci}
Our final example is the rotating AdS$_4$ black hole solution with no twist \cite{Hristov:2019mqp}, such that $\sum_I p^I = 0$. There are again two fixed points at the centre of AdS$_2$ and the two poles of the sphere in their near-horizon region. The gluing rule was determined in \cite{Hosseini:2019iad}: both fixed points exhibit $s=1$ and the same arbitrary deformation $\omega$ with $X^I = \chi^I \mp \omega p^I$, respectively, and the overall constraint $g_I \chi^I = \sum_I \chi^I = 1 + \omega$. After the conversion to field theory variables $\chi^I = \frac12 \Delta_i$, $p^I = - \frac12 \frak{n}_i$, we arrive at \eqref{eq:sci} with \eqref{eq:BC} under the assumption \eqref{eq:csgn} and the related subtlety. The one-point functions are again non-vanishing and subject to the $\cI$-extremization as in \eqref{eq:Iextrem}, \footnote{In this case the constraint between $\chi^I$ and $\omega$ translates into a more complicated constraint involving both electric charges $\frak{q}_i$ and angular momentum $\frak{J}$, see \cite{Choi:2018fdc}.}. Unfortunately we have no available results in the literature regarding the finite $N$ evaluation of the SCI for ABJM. We turn to discuss two special limits that make the field theory object more tractable.

{\bf Vanishing magnetic charges.} In this case we are free to completely turn off the magnetic charges, which simplifies the SCI to
\bea
	&& Z_\text{SCI} \simeq \text{Ai}^2 (C^{-1/3} (N_k - B))\ , \\
\nonumber	C = && \frac{2 \omega^2}{\pi^2 k \prod_i \Delta_i}\ ,  B = \frac{(1+\omega)^2 k_\mathbb{T} (\Delta) + (1-\omega)^2 k_\mathbb{W} (\Delta)}{48 k \prod_i \Delta_i}\ ,
\eea
under the constraint $\sum_i \Delta_i = 2 (1+\omega)$.

{\bf Cardy limit.} A further simplification takes place when we take $\omega \rightarrow 0$ ($\omega = 0$ is degenerate since one cannot turn off the rotation here). In this case we find
\be
	-\log Z_{SCI}^\text{Cardy} \simeq\frac{2 \pi \sqrt{2 k \Delta_1 \Delta_2 \Delta_3 \Delta_4}}{3 \omega}\, N_{k, \Delta}^{3/2} + \frac12 \log  N_{k, \Delta}\ ,
\ee
under the constraint $\sum_i \Delta_i = 2$, matching (upto an overall factor of $2$) with \eqref{eq:cardy}. This limit was investigated in \cite{Choi:2019dfu,Nian:2019pxj} at large $N$, in agreement with the above result.

\section{Outlook}
\label{sec:out}
A number of questions, apart from the imminent task of proving the conjecture in \cite{Hristov:2021qsw}, remain open. 

So far we omitted from the discussion the constant (in $N$) and non-perturbative corrections, even if these are to a certain extent already known for the ABJM squashed sphere partition function. These additional terms have no obvious origin in the gravitational blocks \eqref{eq:gravNek} and must therefore come either from additional gravitational saddles or entirely from quantum corrections. We could then similarly include them inside the full quantum-gravitational blocks \eqref{eq:qgravNek} and apply the gluing rules, but due to the lack of proper understanding at this stage this is a purely formal exercise. It is instead desirable to derive all quantum corrections (perturbative and non-perturbative) from first principles, which in the current approach amounts to rigorously defining the Nekrasov partition function \cite{Nekrasov:2002qd} in the presence of the additional supergravity multiplet. 

Another important task is the independent verification of the higher derivative supergravity theory defined by \eqref{eq:central}-\eqref{eq:identif} as a truncation from $\cN=8$ supergravity and in turn from an explicit compactification of 11d supergravity on S$^7$.

\section*{Acknowledgements}
I am grateful to Seyed Morteza Hosseini and Valentin Reys for useful discussions and healthy skepticism. I am supported in part by the Bulgarian NSF grants N28/5 and KP-06-N 38/11.

\bibliographystyle{apsrev4-2}
\bibliography{ABJM.bib}

\end{document}